# Analysis of a Reputation System for Mobile Ad-Hoc Networks with Liars

Jochen Mundinger
EPFL-IC-LCA
BC203, Station 14
CH-1015 Lausanne, Switzerland
jochen.mundinger@epfl.ch

Jean-Yves Le Boudec
EPFL-IC-LCA
BC203, Station 14
CH-1015 Lausanne, Switzerland
jean-yves.leboudec@epfl.ch

October 29, 2018

**Abstract**

The application of decentralized reputation systems is a promising approach to ensure cooperation and fairness, as well as to address random failures and malicious attacks in Mobile Ad-Hoc Networks. However, they are potentially vulnerable to liars. With our work, we provide a first step to analyzing robustness of a reputation system based on a deviation test. Using a mean-field approach to our stochastic process model, we show that liars have no impact unless their number exceeds a certain threshold (phase transition). We give precise formulae for the critical values and thus provide guidelines for an optimal choice of parameters.

## 1 Introduction

The performance of Mobile Ad-Hoc Networks is well-known to suffer from *free-riding* as there is a natural incentive for nodes to only consume, but not contribute to the services of the system. Forwarding their peers' packets, for example, uses up limited battery lifetime without giving any immediate reward. Users themselves do not need to be technology experts; it would suffice if a company produced devices that do not cooperate and therefore have a longer runtime.

Free-riding is a well-known problem in economics and often occurs in situations involving *public goods* [12]. That is, goods that are both non-rivalrous (meaning that consumption by one does not limit consumption by others) and non-exclusive (meaning benefits cannot be excluded). In Mobile Ad-Hoc Networks, nodes free-ride in that they increase their utility by not shouldering their fair share of the costs.

On the other hand, altruism has been observed. *Altruism* is the practice of being helpful to other people with little or no interest in being rewarded for one's efforts. The concept has a long history in philosophical and ethical thought. For recent references see [41, 19, 4]. However, we cannot typically rely on altruistic nodes to ensure collaboration in Mobile Ad-Hoc Networks. Moreover, we would like to protect the network from malicious attacks and random failures. In the absence of suitable countermeasures, performance of the network can dramatically decrease [31, 8, 1].

A solution approach addressing all of these problems is the application of reputation systems. By *reputation* we mean an estimate about a nodes's actual quality in terms of its behaviour in the network. Sometimes this is also referred to as *trust*. Nodes keep track of their peers' behaviour and exchange this information with others in order to compute a reputation value about the other nodes. Nodes with a good reputation are then favoured. For more general articles on reputation systems see [42] and [32]. However, in the context of Mobile Ad-Hoc Networks a reputation system must necessarily be fully decentralized to fit the architecture and to preserve its advantages such as scalability.

By using second-hand information in addition to own observations, an accurate estimate of some *subject's* (the node under consideration) behaviour can be obtained faster. Moreover, a node can have a reputation value about a subject without ever having interacted with it itself. However, an inherent problem with any such mechanism is the vulnerability to liars.

A simple idea to protect the system from liars was suggested by Buchegger and Le Boudec as part of their CONFIDANT Protocol (Cooperation of Nodes, Fairness In Dynamic Ad-hoc NeTworks) [8, 9]. A node believes second-hand information only if it does not differ too much from the node's reputation value. This is called the deviation test. In fact, the system considered in [9] is more complex. It also allows for using second-hand information from trusted peers, where here trust refers to agreement in the reputation values. Therefore each node maintains both a reputation and a trust value about each of its peers.

The system appears to work well, although performance has only been evaluated through simulations of a network with a particular set of assumptions (e.g. on the routing protocol). Further simulations suggested that the deviation test on its own without the trust component performs nearly as well. It seems surprising that such a simple idea works so well and we consider it worth analyzing in more detail and in a more general context. This is the aim of our research.

We thus consider an abstract model of a reputation system based on the deviation test. We are primarily concerned with the formation of reputation rather than with the detection and response components of a reputation system. The detection component depends on the given Mobile Ad-Hoc Network. For example, most reputation systems suggested for Mobile Ad-Hoc Networks have been assuming Dynamic Source Routing (DSR) [24] and have based the detection component on its properties. In contrast, we merely assume that bad behaviour can be told apart from good behaviour. We also assume that if reputation values can be computed accurately, than there exists a response mechanism using them to obtain the desired effects. Typically, this might mean exclusion of benefits for the misbehaving node.

In an earlier paper [39] we dealt with a simplified one-dimensional model. We now consider the original two-dimensional system. This paper is an extended version of the conference paper [38]. Whereas in the one-dimensional model there is only a notion of good and bad reputation, in the two-dimensional model there is an additional notion of the degree of certainty about the reputation. We analyze the case of 2 nodes, one honest and the other a liar. Our new results confirm the results of the simplified one-dimensional model.

Our model is also relevant in the context of social networks. If a person in the network is confronted with information that is not verifiable, they will probably believe it only if, to them, it seems likely. However, they will ignore it if, to them, it seems unlikely. Work in this context can be found in [40].

The methodology we have developed for this problem is a follows. We first formulate a stochastic process that models the system based on certain assumptions. We then derive the 'mean' ordinary differential equation (ODE) by averaging the dynamics and passing to a *fast-time scaling* limit. We solve the ODE and study its fixed points. Our approach is thus a *mean-field* approach. Full details are given in Sections 4 and 5. Finally, we verify the analytical results by means of simulation.

We find that there is a *phase transition*. That is, there is a threshold proportion of lying below which the lying has no impact. Above it, the lying does have an impact and corrupts the reputation system. We give precise formulae and quantify the impact, thereby providing a performance evaluation of a reputation system based on the deviation test. Alternatively, this phase transition can be phrased in terms of the system parameter controlling the deviation test. We thus provide guidelines for a good choice of parameters and hence a good system design.

This paper is organized as follows. We present related work in Section 2. The precise modelling assumptions are listed in Section 3 and the stochastic process model is formulated in Section 4. We give details of our methodology and provide mean-field results in Section 5 and verify them by means of simulation in Section 6. We conclude in Section 7.

## 2  Related Work

A number of reputation mechanisms have been suggested and studied. A comprehensive survey and more detailed overview of reputation systems for Mobile Ad-Hoc Networks can be found in [10]. Reputation systems are classified according to (1) representation of information and classification, (2) use of second-hand information, (3) trust and (4) redemption and secondary response.

In addition to the CONFIDANT protocol already discussed in the introduction, the COllaborative REputation mechanism (CORE) was introduced in [34]. Here, reputation also takes into account a task-specific functional reputation. OCEAN [7] and SORI [23] are also discussed in more detail in the survey. These systems have all been developed for a fairly specific set of assumptions, in particular assuming Dynamic Source Routing (DSR) [24].

Reputation systems have also been considered in other scenarios, in particular Internet-based Peer-to-Peer (P2P) systems. Aberer and Despotovic [2] suggest a mechanism for P-Grid, a P2P system, that spreads negative information only. Collaboration enforcement in P2P systems has also been considered by Moreton and Twigg [37]. Another mechanism is PeerTrust as introduced by [48]. The reader is referred to [25] for the EigenTrust algorithm, a method to compute global trust values in the presence of pre-trusted peers. Another mechanism is PeerTrust as introduced by [48]. A comprehensive survey and more detailed overview of reputation systems for Internet-based P2P systems can be found in [3]. Another review with a focus on the artificial intelligence literature is given in [43].

Our work is the first analytical approach to evaluate robustness against liars. [50] consider the problem of liars via some models of deception. Their approach is based on the weighted majority technique where the last second-hand information is tested by comparing it to the next direct interaction. The analysis is based only on simulation. [47] is also concerned with filtering out manipulated second-hand information that seems unlikely. However, they consider quantiles of the Beta distribution rather than distance. This paper, too, is based merely on simulation. In the context of centralized reputation systems, [35] consider incentive mechanisms not in order to stimulate good behaviour in the network, but to stimulate honest reports within the reputation system. They show that honest reporting is a Nash equilibrium.

Alternatives to reputation systems have also been considered, e.g. incentive mechanisms. They make it advantageous for nodes to act in such a way that the resulting social welfare is optimal (e.g., [6]). In general, this can either be achieved by means of pricing schemes [13] or rules [5]. As for pricing schemes, they might involve payments in kind, or virtual or real payments. For example, a virtual currency called nugglets is suggested in [11] and a pricing scheme based on notional credit is used in [15], both in the context of Mobile Ad-Hoc Networks. In [22], the authors examine a micro-payment mechanism for P2P file sharing networks. A simple example of a rule is to force nodes to contribute in order to consume. Rules are used in [14] in the context of P2P Wireless LAN Consortia. Incentive mechanisms typically give rise to game-theoretic problems. Unlike reputation systems, incentive mechanisms do not address malicious nodes and random failures.

A more recent alternative to reputation systems is artificial immune systems. They are aimed primarily at misbehaviour detection and designed so that they adapt to normal behaviour, but they also recognize new misbehaviour patterns that were not anticipated in the system design phase [45]. Moreover, artificial immune systems use mechanisms for faster detection of repeated misbehaviour [28] and [44]. An important potential advantage of such systems is their inherent randomness that provides diversity at the population level. Even if some computers are vulnerable to an attack, there should be many others that are resistant to the same attack. However, implementations seem to depend very much on the particular application and at the moment there is no scenario in which artificial immune system have proven superior to other approaches.

All three approaches – incentive mechanisms, reputation systems and artificial immune systems – need to take into account both the economic side and the engineering side. For example, identity is an issue in all these systems. For more information, the reader is referred to [21, 18].

## 3 Assumptions

### 3.1 Subject Behaviour

We consider the case when there is a single subject whose reputation is considered. This subject might be one of the $N$ nodes themselves, but it might also be the provider of some external service such as Internet access. At each observation, its actual *behaviour* is assumed to be either positive or negative with probabilities $\theta$ and $1 - \theta$ respectively, independently of all other observations. The more practical case when there are $M$ subjects of interest can be decomposed into $M$ instances of our model, as the $M$ different sets of reputation values do not interfere with each other. In particular, this allows all nodes to be subjects themselves, as is the case in a Mobile Ad-Hoc Network.

### 3.2 Reputation

Each node $i$ maintains a *reputation value* $R^i = R_n^i$ about the subject that reflects its belief about $\theta$ at the time of the $n$th interaction. The initial value is $R_0^i$. Reputation values are based on two counters, as will be explained in Section 4. This opinion might change with new observations, arising either from interactions with the subject itself or with a peer.

A *direct (first hand) observation* is an observation of the subject's behaviour. Direct observations are always accepted and the reputation values updated accordingly. An *indirect (second-hand) observation* arises from interactions with peers who report about their own observations. Indirect observations are only accepted if the reported observation does not deviate too far from the current reputation $R_n^i$. This deviation test is controlled by the parameter $d$. Moreover, if accepted, the impact of an indirect observation is scaled by a weighting parameter $\omega$. Finally, we account for discount factor $u$ so that the system gradually forgets about old observations. This also allows for tracking subject behaviour that changes.

We shall restrict attention to the case of 2 nodes, an honest one with reputation value $R_n$ about the subject and the other a liar. This is closely related to the general case, because several liars can be thought of as a single liar by aggregating their influence and because we can focus on one out of the honest nodes by symmetry. Furthermore, it looks like ignoring the other honest ones can be accounted for by increasing the proportion of (necessarily truthful) direct interactions, but this will have to be investigated in more detail (cf. Section 7).

## 3.3 Interaction Model

In the general case, the nodes make observations when interacting with the subject or with a peer. Interactions depend on the topology of the network and the mobility model for the nodes. We take a high-level view and directly specify the interactions rather than topology and mobility. More specifically, we assume that interactions occur at certain points $T_n$ in time and that a given resulting observation is direct or indirect with fixed probabilities $p$ and $\bar{p} = 1 - p$ respectively. As a natural model, let the $T_n$ be the points of a Poisson process and consecutive interactions be independent. Without loss of generality we take the Poisson process to have rate 1, otherwise we can scale time suitably. Whereas for a given Mobile Ad-Hoc Network the values of $p$ and $\bar{p}$ are determined by topology and mobility, we examine the whole range of parameters $p = 1 - \bar{p} \in [0, 1]$. $p$ large accounts for Mobile Ad-Hoc Networks where direct observations are frequent, $p$ small accounts for networks where they are rare.

For ease of notation, we shall often refer to the number of the observation $n$ rather than its point in time $T_n$.

Finally, note that in the case of 2 nodes with one honest node and one liar, the honest node's second-hand information is necessarily from its lying peer.

## 3.4 Adversary Model

One needs to make precise assumptions about the adversaries' abilities in order to give performance guarantees. We assume that liars follow the plain strategy to always lie maximally, i.e. they will always report either extremely negative or extremely positive behaviour about the subject when interacting with their peers in an attempt to achieve maximal impact. It suffices to focus on the extremely negative part, as the other one is similar by symmetry.

## 3.5 Performance

Notice that liars can easily change their own reputation values to anything they want. The question is whether they can influence the reputation values of the honest nodes.

The *faster* the nodes can obtain *accurate* estimates of the subject behaviour, the better the system will work, but there is a fundamental trade-off between *speed* and *robustness*. By using more second-hand information, an estimate of the subject's behaviour can be obtained faster – a node can even have a reputation value about a subject without ever having interacted with it itself; however, it is then more sensitive to potentially wrong information and one needs to compromise on accuracy.

We assess robustness in detail. It will then be possible, for example, to choose parameters such that the system will be as fast as possible, subject to a given accuracy condition. In fact, we show that by using a reputation system based on the deviation test, to some extent it is possible to gain in terms of speed without compromising on accuracy.

## 4   Model

We first formulate a stochastic process that models the systems based on the assumptions of the previous section. A natural scheme, suggested by the reputation system in [9] and other proposals, is to keep a history of previous events. Two counters, $\alpha_n$ and $\beta_n$, are updated whenever there is a new observation, either direct or indirect. $\alpha_n$ keeps track of positive observations, $\beta_n$ keeps track of negative observations. Thus we are led to consider the following two-dimensional stochastic process $(\alpha_n, \beta_n)$ for $n \geq 0$.

$$(\alpha_{n+1}, \beta_{n+1}) = u(\alpha_n, \beta_n) + \begin{cases} (1,0) & \text{w.p. } p\theta \\ (0,1) & \text{w.p. } p(1-\theta) \\ (0,\omega)\mathbf{1}\left\{\frac{\alpha_n}{\alpha_n+\beta_n} \leq d\right\} & \text{w.p. } \bar{p} \end{cases} \quad (1)$$

The three possibilities correspond to a positive direct observation, a negative direct observation and an indirect observation respectively (cf. Sections 3.1 and 3.3). Direct observations are always accepted and counted with 1. Indirect observations have to pass the deviation test with parameter $0 < d < 1$, modelled by the indicator function, and are weighted by $\omega > 0$ as described in Section 3.2. Indirect observations are negative, because they are obtained from the liar who is assumed to report extremely negative behaviour (cf. Section 3.4). In any case, both components are discounted individually by $0 < u < 1$ (cf. Section 3.2). Typically, $u$ is close to 1.

The stochastic process (1) is a homogeneous Markov Chain with the state space being a subset of the triangular area $\{(\alpha, \beta) : \max\{1, \omega\}\alpha + \beta \leq \max\{1, \omega\}/(1-u)\}$ in the first quadrant. This is illustrated in Figure 1. For the parameters chosen to be rational, the process will take rational values only and the state space is easily seen to be countable, although not easy to describe.

The important quantity of interest is $R_n = \alpha_n/(\alpha_n + \beta_n)$, in some sense the proportion of positive observations. We examine how well this compares to the true subject behaviour $\theta$. In addition, in this two-dimensional model, there is the notion of the degree of certainty about reputation values, determined by $\alpha_n + \beta_n$. The larger this sum, the more certain the node is of its reputation value, the more 'locked in' the state. Lines of constant reputation, as well as constant certainty, are also shown in Figure 1.

The initial values are $(\alpha_0, \beta_0)$. Note that if $\omega = 1$ and the deviation test is passed, then starting with $\alpha_0 + \beta_0 = 1/(1-u)$ will leave $\alpha_n + \beta_n$ unchanged; starting elsewhere the sum will converge to $1/(1-u)$. It makes sense to start with such a converged value, because we would like to allow for tracking behaviour that changes. However, there can be no a priori knowledge of a change, so we cannot simply reset the system to an arbitrary starting value. On the other hand, as the fixed points are not known in advance, we cannot generally choose initial values to correspond to them. We still take initial values satisfying $\alpha_0 + \beta_0 = 1/(1-u)$ as this choice can be viewed as an extreme case. Moreover, we shall see in Sections 5 and 6 that, the choice of initial value does not influence the process in the long run. If we observe the process settling down at a fixed point starting from such an initial state, then it will also do so from a less 'locked in' state with the same reputation value. By the initial state $R_0$ we shall mean $\frac{1}{1-u}(R_0, 1 - R_0)$. Notation is summarized in Table 1.

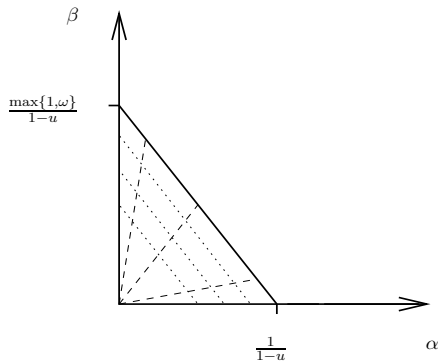

Figure 1: **The state space of the stochastic process (1) is a subset of the triangular area. Dashed lines show points of constant reputation, increasing from bottom to top. Dotted lines show points of constant certainty, increasing from left to right.**

| symbol | meaning |
|---|---|
| $\theta$ | prob. of positive subject behaviour (cf. Section 3.1) |
| $p$ | prob. of an obs. being direct (cf. Section 3.3) |
| $\bar{p}$ | prob. of an obs. being indirect (cf. Section 3.3) |
| $\alpha_n$ | positive reputation component (cf. Section 4) |
| $\beta_n$ | negative reputation component (cf. Section 4) |
| $R_0$ | initial reputation value (cf. Sections 3.2, 4) |
| $R_n$ | inferred reputation value (cf. Sections 3.2, 4) |
| $d$ | deviation test parameter (cf. Section 3.2) |
| $\omega$ | weighting factor for indirect obs. (cf. Section 3.2) |
| $u$ | discount factor (cf. Section 4) |

Table 1: **Summary of notation**

Note that, although we have defined the process in order to estimate $\theta$, it does not converge to a constant (in probability). For all $n$, there is positive probability of the next state taking either one of two values that differ by a constant. This is due to the discounting. So, we assess convergence (in distribution) to some limiting distribution from which we infer $\theta$. We will point out another advantage of the discounting in the reputation system that is essentially due this convergence in distribution.

# 5 Mean-field Approach

Our methodology is as follows. From the stochastic process formulation (1), we derive the 'mean' ordinary differential equation (ODE) by averaging the dynamics and passing to a *fast-time scaling* limit. That is, we scale time so that events occur more frequently, that is the honest node makes direct and indirect observations at a higher rate. At the same time, the impact of each observation is reduced by the same factor. A similar approach has also been used in TCP modelling (cf. [36] and [30]). One possibility to turn this type of argument into a rigourous formal limit statement is the fluid limit in [16]. We then derive the solutions of the differential equation and study its fixed points. We confirm the analytical results by means of simulation. This is the topic of the next section.

Thus, our approach can be called a *mean-field* approach. This term has different interpretations in different research communities. In [49], it refers to making an independence assumption for one side of the Chapman-Kolmogorov equations. The mean-field idea was first introduced in physics and has been used to describe systems like plasma and dense gases where interaction between particles is somewhat weak, meaning that the strength of the interaction is inversely proportional to the size of the system. A given particle is seen as under a collective force generated by the other particles. In recent years, the mean-field approach has also found an appeal in the area of communication networks. To name but a few of the many publications, the reader is referred to [27], [46], [26] and [17]. The stochastic approximation framework by [29] is also a mean-field approach. The basic paradigm is a stochastic difference equation where one recursively adjusts the parameter so that some goal is met asymptotically. The main concept used is to show that noise effects average out so that the process is determined by a mean ODE. This has been applied in diverse areas, in particular in signal processing and communications.

The mean-field approach is very powerful in that it enables us to study a complex system analytically without losing the important features. As we shall see, the phase transition, as well as the fixed points and critical parameter values, is predicted accurately. What it cannot do, however, is to compare two simultaneous fixed points in terms of their attraction. It does not keep all the information contained in the stochastic process. For this, we would need to investigate a diffusion approximation approach as in [20]. We will gain some insight into this question via the simulations.

Specifically, we consider a family of processes indexed by $N$. We will now make the notation explicit again to stress that the subscripts $n$ are the points $T_n$ of a Poisson process at rate 1.

$$(\alpha^N_{T_{n+1}/N}, \beta^N_{T_{n+1}/N}) = (1 - \frac{1-u}{N})(\alpha^N_{T_n/N}, \beta^N_{T_n/N})$$
$$+ \frac{1}{N}\begin{cases} (1,0) & \text{w.p. } p\theta \\ (0,1) & \text{w.p. } p(1-\theta) \quad (2) \\ (0,\omega)\mathbf{1}\left\{\alpha^N_{T_n/N}/(\alpha^N_{T_n/N} + \beta^N_{T_n/N}) \le d\right\} & \text{w.p. } \bar{p} \end{cases}$$

Next, we consider a continuous-time rescaled version. The number of jumps of the process indexed by $N$ in the interval $[t, t+\epsilon)$ is Poisson($N\epsilon$), the average jump is of size

$$\begin{matrix} -\frac{1-u}{N}\alpha^N_t + \frac{1}{N}\left[p\theta\right] \\ -\frac{1-u}{N}\beta^N_t + \frac{1}{N}\left[p(1-\theta) + \bar{p}\omega\mathbf{1}\left\{\frac{\alpha^N_t}{\alpha^N_t+\beta^N_t} \le d\right\}\right] \end{matrix} \quad (3)$$

We obtain

$$\begin{aligned}
\alpha^N(t+\epsilon) - \alpha^N(t) &= \tfrac{N\epsilon}{N}\left[(u-1)\alpha(t) + p\theta\right] \\
\beta^N(t+\epsilon) - \beta^N(t) &= \tfrac{N\epsilon}{N}\left[(u-1)\beta(t) + p(1-\theta) + \omega\bar{p}\mathbf{1}\left\{\tfrac{\alpha}{\alpha+\beta} \leq d\right\}\right]
\end{aligned} \quad (4)$$

Dividing by $\epsilon$ and taking the limit, we are thus led to consider the following deterministic ODE.

$$\begin{aligned}
\dot{\alpha}(t) &= (u-1)\alpha(t) + p\theta \\
\dot{\beta}(t) &= (u-1)\beta(t) + p(1-\theta) + \omega\bar{p}\mathbf{1}\left\{\tfrac{\alpha}{\alpha+\beta} \leq d\right\}
\end{aligned} \quad (5)$$

This system is discontinuous, but linear above and below the line of discontinuity $\alpha/(\alpha+\beta) = d$. Above it, for $\alpha/(\alpha+\beta) > d$, we obtain

$$\begin{aligned}
\alpha(t) &= c_1 e^{-(1-u)t} + \tfrac{1}{1-u}p\theta \\
\beta(t) &= c_2 e^{-(1-u)t} + \tfrac{1}{1-u}p(1-\theta)
\end{aligned} \quad (6)$$

for constants $c_1, c_2$ whereas below it, for $\alpha/(\alpha+\beta) \leq d$, we obtain

$$\begin{aligned}
\alpha(t) &= c_3 e^{-(1-u)t} + \tfrac{1}{1-u}p\theta \\
\beta(t) &= c_4 e^{-(1-u)t} + \tfrac{1}{1-u}[p(1-\theta) + \omega\bar{p}]
\end{aligned} \quad (7)$$

for constants $c_3, c_4$. Fixing the initial reputation counter at $(\alpha_0, \beta_0)$ we obtain the constants and hence the solutions. We find that the system has either one or two fixed points, depending on the parameters of the model.

$$(\alpha, \beta) = \frac{p}{1-u}(\theta, 1-\theta) \quad (8)$$

is a fixed point is a fixed point if $\theta > d$. If it exists, it is asymptotically stable and trajectories from $\alpha/(\alpha+\beta) > d$ are attracted to it. The corresponding reputation value is $\theta$.

$$(\alpha, \beta) = \left(\frac{p\theta}{1-u}, \frac{p(1-\theta)}{1-u} + \frac{\omega\bar{p}}{1-u}\right) \quad (9)$$

is a fixed point if $\theta \leq d$ or $\bar{p} \geq (\theta - d)/(\theta - d + \omega d)$. If it exists, it is asymptotically stable and trajectories from $\alpha/(\alpha+\beta) \leq d$ are attracted to it. The corresponding reputation value is $\pi = \theta p/(p + \omega\bar{p})$. If only one of the two fixed points exists then the trajectories from the other region lead into its region and thus are also attracted to it. That is all trajectories are attracted to it. Otherwise, both are asymptotically stable on their respective region. Hence, we have the following result.

**Theorem 1** *If $\theta > d$, (8) is a fixed point of the mean ODE (5). For $\bar{p} < \bar{p}_c = (\theta - d)/(\theta - d + \omega d)$ it is asymptotically stable and all trajectories are attracted to it. The corresponding reputation value is the true $\theta$. Otherwise, there is a second, false fixed point (9) and both are asymptotically stable, attracting trajectories from $\alpha/(\alpha+\beta) > d$ and $\alpha/(\alpha+\beta) \leq d$ respectively. If $\theta \leq d$ then only the latter, false one is asymptotically stable and all trajectories are attracted to it. The corresponding reputation value is $\pi = (p\theta)/(p + \omega\bar{p})$.*

Thus, the reputation system exhibits a phase transition. Assuming $\theta > d$, we find a bifurcation in terms of the parameter $\bar{p}$. In the *subcritical regime*, that is, for $\bar{p} < \bar{p}_c$, the fixed point corresponding to the true $\theta$ is unique. In the *supercritical regime* where $\bar{p} \geq \bar{p}_c$ there is a second, false fixed point.

In practical terms, this suggests that the reputation system works and that the liar cannot achieve anything if $\theta > d$ and $\bar{p} < \bar{p}_c$. However, the liar does have an impact otherwise. As for the latter condition, it is intuitively clear that the deviation test can filter out extreme lies only if they do not occur too often. As for the first condition, it is clear that if the true $\theta$ is too close to the extreme 0 behaviour, the deviation test will not filter the lies and the liar will have an impact. The deviation test cannot protect a 'very bad' subject behaviour to be pushed by the liar to an 'extremely bad' perception by the honest node. However, there is a range of parameters for which it does protect the reputation system.

As mentioned in Section 3, we can repeat the analysis to show that the reputation system similarly protects against extremely positive reports. Combining the two, we obtain the following necessary and sufficient conditions for the true fixed point to be unique when both, positive and negative lying are permitted: $\min\{\theta, 1 - \theta\} > d$ and $\bar{p} < (\min\{\theta, 1 - \theta\} - d)/(\min\{\theta, 1 - \theta\} - d + \omega d)$.

Alternatively, the phase transition can be phrased in terms of the system parameter $d$. For small $d$ there is only one fixed point, the true one (8). For intermediate $d$, there are both the true and the false fixed points and for large $d$ there is only the false fixed point (9). The exact conditions are given in Corollary 1 and an illustration is provided in Figure 2.

**Corollary 1** *If $d < d_{c_1} = \pi = (p\theta)/(p + \omega\bar{p})$, (8) is the unique fixed point of the mean ODE (5). It is asymptotically stable and all trajectories are attracted to it. The corresponding reputation value is the true theta. If $d_{c_1} \leq d < d_{c_2} = \theta$ there is a second, false fixed point (9). Both are asymptotically stable, attracting trajectories from $\alpha/(\alpha + \beta) > d$ and $\alpha/(\alpha + \beta) \leq d$ respectively. If $d_{c_2} \leq d$, then only the latter, false one is asymptotically stable and all trajectories are attracted to it. The corresponding reputation value is $\pi = (p\theta)/(p + \omega\bar{p})$.*

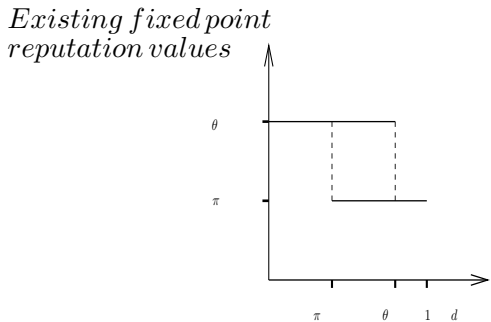

**Figure 2: Bifurcation plot showing the existence of fixed points for each $d$: As $d$ increases from 0 to 1 (from left to right along the horizontal axis), the number of fixed points changes from 1 (the true fixed point at reputation value $\theta$) to 2 (the true fixed point at $\theta$ and the false fixed point at $\pi$) and back to 1 (the false fixed point at $\pi$).**

In practical terms, this suggests that the reputation system works as long as $d$ is sufficiently small. It might not work otherwise, however, and if $d$ is sufficiently large it will not work at all. Although this qualitative inverse dependence between $d$ and robustness against liars is entirely as one would expect, the exact dependence might seem surprising. It can be exploited for system optimization as follows. If we are interested in a system that is as fast a possible subject to being accurate, we would choose $d$ just below the first critical value $\pi$. Then this is the largest value of $d$ for which the true $\theta$ is still the unique fixed point reputation value. As such it allows for maximal gain in terms of speed using second-hand information while still being accurate. This will protect the reputation systems from liars for a reasonable range of parameters. Given a more general cost function with arbitrary weights on accuracy and speed, we could again compute the optimal choice of the system parameter $d$. Thus, we provide guidelines for a good choice of parameters and hence system design.

## 6 Simulations

In this section we report our simulation results. We used formulation (1) to compute $10^5$ steps and then plot $R_n = \alpha_n/(\alpha_n + \beta_n)$ against $n$. The program code was written in Java (Version 1.4.1). We used both the standard Java random number generator and the Mersenne Twister (MT19937) [33] to generate pseudo-random numbers. The lower and upper boundaries in the graphs correspond to reputation values 0 and 1 respectively. The lower and upper intermediate lines correspond to the possible fixed point reputation values $\pi = (p\theta)/(p + \omega\bar{p})$ and $\theta$ respectively. 100 independent runs were carried out in each case and a typical sample path is shown. We consider the following set of parameters and variations thereof.

**Parameter set 1** $\theta = 0.8$, $d = 0.4$, $u = 0.99$ and $\omega = 1$ *for various values of $p$ and $\bar{p} = 1 - p$. We used both the extreme initial values $R_0 = 0$ and $R_0 = 1$. Thus, from the previous section, the predicted possible fixed point reputation values are $0.8$ and $0.8p$. The critical lying probability is $\bar{p}_c = p_c = 0.5$.*

In Figure 3 we show a typical sample path for parameter set 1 with $\bar{p} = 0.2$, i.e. $p = 0.8$, and $R_0 = 0$. We note that $R_n$ increases from 0 past $\pi$ to $\theta$ and then remains within its neighbourhood until the end of the simulation. All 100 independent runs showed the same qualitative sample paths. As $\bar{p} = 0.2 < \bar{p}_c$, this is a subcritical scenario. As expected, $\theta$ is confirmed as the unique fixed point reputation value. Indeed, we obtained the same results when $\bar{p} = 0.2$ is replaced by $\bar{p} = 0.4$ and $\bar{p} = 0.45 < \bar{p}_c$. Here, starting with $R_0 = 0$ can be viewed as a worst case. For other starting values, too, including the other extreme $R_0 = 1$, we obtained the same qualitative results.

In Figure 4 we illustrate the effect of the discount parameter $u$. A typical sample path is shown for parameter set 1 except $u = 0.999$. The variability around $\theta$ is smaller and it takes longer for the process to approach $\theta$.

In Figure 5 we consider the supercritical case with now $\bar{p} = 0.8$. Also, unlike in parameter set 1, we take $u = 0.95$ for clearer illustration. As a side effect, variability is increased. Increasing from 0, $R_n$ settles down for some time in the neighbourhoods of $\pi = (p\theta)/(p+\omega\bar{p}) = 0.16$ and $\theta = 0.8$ in an alternating fashion. This is in agreement with the mean-field prediction that both $\pi$ and $\theta$ are fixed point reputation values. It is due to the discounting that we do not have convergence to a constant, but there is always a positive probability of moving from one fixed point to the other. This can be viewed as another advantage of discounting, because the process cannot get stuck at the false fixed point forever. The same qualitative results are observed in all 100 independent runs and also for $\bar{p} = 0.6$ and $\bar{p} = 0.55 > \bar{p}_c$, only the proportion of time spent near $\theta$ is higher. Note also that the false fixed point reputation value depends on $\bar{p}$ and $\omega$: $\pi = 0.32$ and $\pi = 0.36$ respectively.

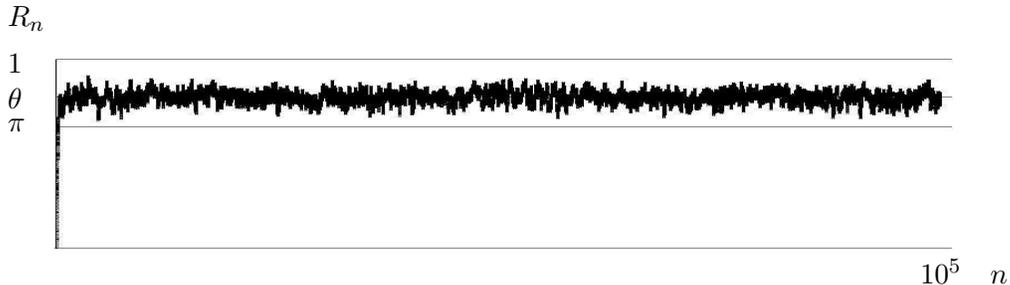

**Figure 3:** $10^5$ steps of a typical sample path for parameter set 1 with $\bar{p} = 0.2$ and $R_0 = 0$. We plot $R_n$ against $n$, the upper line corresponding to $\theta$ and the lower line to $\pi = (p\theta)/(p + \omega\bar{p})$ (cf. Figure 2). $R_n$ increases from $0$ past $\pi$ to $\theta$ and then remains close to $\theta$. Thus, $\theta$ is confirmed as the unique fixed point reputation value as expected from the analytical results.

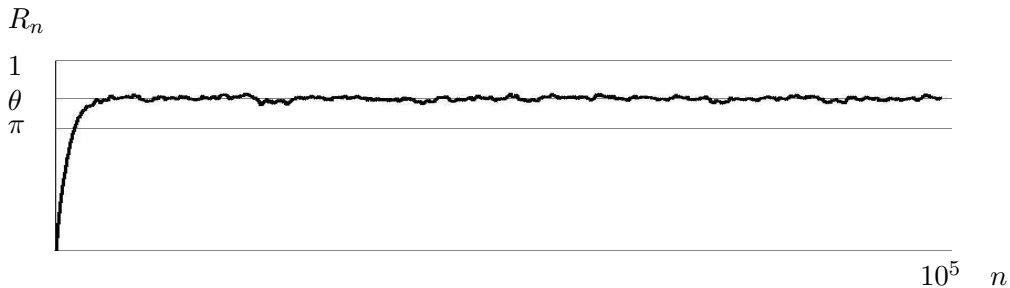

**Figure 4:** $10^5$ steps of a typical sample path for the same parameters as in Figure 3 except $u = 0.999$. The variability is smaller and it takes longer for the process to approach $\theta$.

In Figure 6 we demonstrate that the starting value $R_0$ does not influence the process in the long run. Whereas for the solution of the ODE each fixed point has a basin of attraction (Theorem 1), there is always positive probability for the stochastic process (1) to move from one to the other, so the independence of the starting value is as expected.

With a different choice of parameters, the prediction of only one fixed point at $\pi = (p\theta)/(p + \omega\bar{p})$ for the case $\theta \leq d$ can also be verified. Finally, we carried out simulations with the same parameters as in Figure 3 except $\omega = 2$ to verify, in particular, the critical value $\bar{p}_c = 1/3$.

Thus, we have verified the results of the mean-field approach in Section 5. In addition, Figures 3 – 6 give us an idea of the proportion of time near the false $\pi = (p\theta)/(p + \omega\bar{p})$. This time increases with $\bar{p}$.

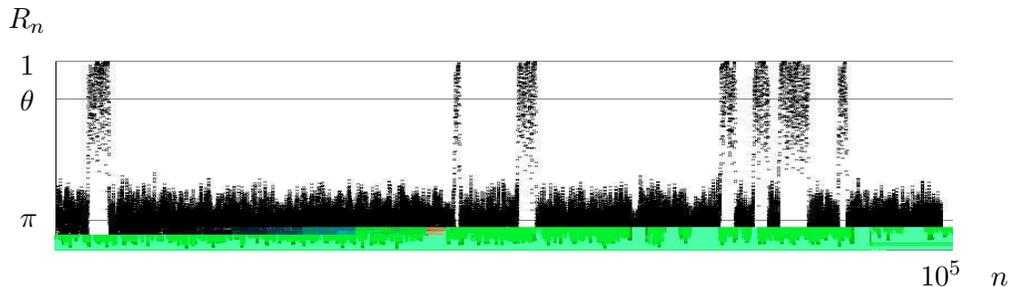

**Figure 5:** $10^5$ steps of a typical sample path for parameter set 1 except $u = 0.95$ with $\bar{p} = 0.8$ and $R_0 = 0$. $R_n$ increases from 0 to $\pi$ and then settles down for some time in a neighbourhood of $\pi = (p\theta)/(p + \omega\bar{p})$ and $\theta$ in an alternating fashion.

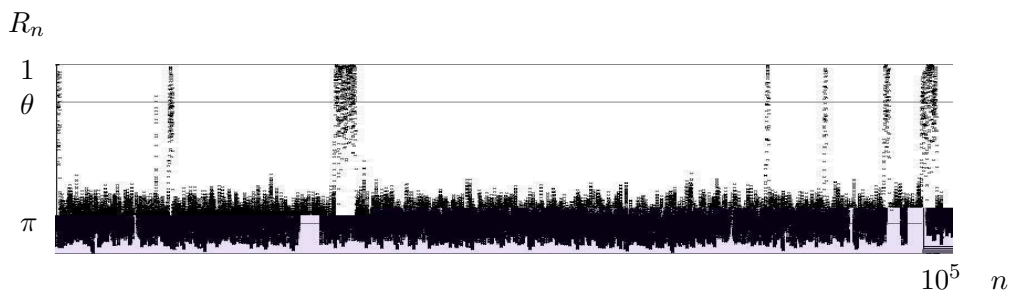

**Figure 6:** $10^5$ steps of a typical sample path for the same parameters as in Figure 5 except $R_0 = 1$. $R_n$ decreases from 1 to $\theta$ and then settles down for some time in a neighbourhood of $\pi = (p\theta)/(p + \omega\bar{p})$ and $\theta$ in an alternating fashion as before (cf. Figure 5). Thus, the starting value $R_0$ does not influence the process in the long run.

## 7 Conclusions and Further Work

The use of decentralized reputation systems is a promising approach to addressing free-riding, as well as random failures and malicious attacks, in Mobile Ad-Hoc Networks. In this paper, we analyze a reputation system based on the deviation test independently of specific implementation assumptions. We consider the case of 2 nodes, one honest and the other a liar. Our results confirm the results obtained for the simplified one-dimensional model addressed in [39].

In regards to the use of second-hand information, there is a fundamental trade-off between speed and robustness. By using second-hand information, an estimate of the subject's behaviour can be obtained faster – a node can even have a reputation value about a subject without ever having interacted with it itself; however, it is then subject to potentially wrong information and one needs to compromise on accuracy. The reputation system based on the deviation test resolves this issue to a certain degree. There is a phase transition in that there is a threshold proportion of lying below which the reputation value of the honest node remains unaffected. Above it, the lying will have an impact and corrupt the reputation system. We give precise formulae and quantify the impact, thereby providing a performance evaluation of the reputation system.

Alternatively, this phase transition can be phrased in terms of the system parameter $d$ controlling the deviation test. We thus provide guidelines for a good choice of parameters and hence system design. For example, for a system that is as fast a possible subject to being accurate, we would choose $d$ just below the first critical value $d_{c1} = \pi$. The deviation test then protects the reputation systems from liars for a reasonable range of parameters.

We have illustrated another fundamental trade-off in terms of the discount parameter. The closer it is to 1, the more accurate the process can estimate the parameter. However, it takes longer to track changing subject behaviour. It is also due to the discount parameter that we do not have convergence to a constant, but there is always a positive probability of moving from one fixed point to another. This can be viewed as another advantage, because the process cannot get stuck at the false fixed point forever.

We have not yet looked in detail at the rate of convergence. Independent runs of the same simulation showed about the same rate of convergence most of the time. We could compare this to the exponential decay terms in the mean ODE and thereby check whether the analytical results are also correct in this respect.

The mean-field approach turned out to be powerful, as well as tractable. We obtained the fixed points as well as the critical points. It cannot, however, compare two simultaneous fixed points in terms of their attraction. For this, we would need to investigate a diffusion approximation approach. Still, even here we gained insight by looking at the time spent near each of the fixed points in the simulation.

The scenario of two nodes that we have considered thus far can also be viewed as an extreme case. Even if all other nodes are malicious so that all second hand information is manipulated, the reputation system protects against the lying in the subcritical regime. In a real-world scenario we would typically be able to assume that at least some if not most nodes in the network are honest, thus indirect observations from honest peers should be considered explicitly.

We have assumed independent subject behaviour. It might be interesting to consider the case when direct observations are correlated.

Another extension is then to consider strategic lying, that is adversaries attempting something more subtle than simply lie maximally. The latter assumption makes it easier to filter out lies with a simple mechanism such as the deviation test, because they are more easily discovered. For example, they could lie in some proportion of reports only or they could always report intermediate behaviour in an attempt to conceal their lies. In this case, as a counter measure, one might also want to think about individually controlled $d^i$, $i = 1, 2, \ldots, N$, based on the nodes' current information. It would also be interesting to consider random noise instead of fake reports.

Finally, it would be of interest to consider a different interaction model, in particular one with a non-homogeneous population of nodes. There might even be some nodes that never interact with the subject directly.

**Acknowledgment**   We would like to thank Sonja Buchegger for valuable discussions.